\documentclass[preprintnumbers,floatfix,aps]{revtex4-1}
\usepackage{amsmath}
\usepackage{amsfonts}
\usepackage{amssymb}
\usepackage{physymb}
\usepackage{graphicx}
\usepackage{epsfig}

\usepackage{dcolumn}
\usepackage{bm}
\usepackage{fancybox}
\usepackage{multirow}
\usepackage{float}
\usepackage{hyperref}
\usepackage{epstopdf}

\usepackage[utf8]{inputenc}

\newcommand{\Punkte}{0}

{\noindent {\bf Exercise {#1}.} #2 \vspace{0.2cm} \\  
\renewcommand{\Punkte}{#3}}%
{\mbox{\hspace{3ex}} \hfill {\bf \Punkte~\mbox{Points}}\bigskip }

\newenvironment{Exercise*}[2]%
{\noindent {\bf Exercise {#1}*.} #2 \vspace{0.2cm} \\ 
renewcommand{\Punkte}{#2}}%
{\mbox{\hspace{2ex}} \hfill {\bf \Punkte~\mbox{Points}}\bigskip }

{\noindent {\bf Exercise {#1}.}\vspace{0.2cm} #2}%
{}

\newcounter{enum1}

\newcounter{enuma}

\begin{document}

\title{{\bf Stability and Tunneling Dynamics of a Dark-Bright Soliton Pair in a Harmonic Trap}}
\date{\today}
\author{E.T. Karamatskos}
\affiliation{Zentrum f\"ur Optische Quantentechnologien,
Universit\"at Hamburg, Luruper Chaussee 149, 22761 Hamburg, Germany}

\author{P. G. Kevrekidis \thanks{%
Email: kevrekid@math.umass.edu}}
\affiliation{Department of Mathematics and Statistics, University of Massachusetts,
Amherst, MA 01003-4515, USA}

\affiliation{Center for Nonlinear Studies and Theoretical Division, Los Alamos
National Laboratory, Los Alamos, NM 87544}

\author{J. Stockhofe}
\affiliation{Zentrum f\"ur Optische Quantentechnologien,
Universit\"at Hamburg, Luruper Chaussee 149, 22761 Hamburg, Germany}

\author{P. Schmelcher}
\affiliation{Zentrum f\"ur Optische Quantentechnologien,
Universit\"at Hamburg, Luruper Chaussee 149, 22761 Hamburg, Germany}
 
\affiliation{The Hamburg Centre for Ultrafast Imaging,
Luruper Chaussee 149, 22761 Hamburg, Germany}

\begin{abstract}
We consider a binary repulsive Bose-Einstein condensate in a harmonic trap in one spatial dimension and investigate particular solutions consisting of two dark-bright (DB) solitons. 
There are two different stationary solutions 
characterized by the phase difference in the bright component, 
in-phase and out-of-phase states. We show that above a critical particle number in the bright component, 
a symmetry breaking bifurcation of the pitchfork type occurs that leads to a 
new asymmetric solution whereas the parental branch, 
i.e., the out-of-phase state becomes unstable. 
These three different states support different small amplitude oscillations,
characterized by an almost stationary density of the dark component and a tunneling of the bright component between the two dark solitons.
Within a suitable effective double-well picture, these can be understood as
the characteristic features of a Bosonic Josephson Junction (BJJ),
and we show within a two-mode approach that all characteristic features of the BJJ phase space are recovered.
For larger deviations from the stationary states, the simplifying double-well description breaks down due to the 
feedback of the bright component onto the dark one, 
causing the solitons to move. 
In this regime we observe intricate anharmonic and aperiodic dynamics, 
exhibiting  remnants of the BJJ phase space.
\end{abstract}

\maketitle

\newpage
\section{Introduction}

Over the past two decades, atomic Bose-Einstein condensates (BECs) have
constituted a pristine platform which has enabled numerous theoretical
and experimental investigations; see e.g.~\cite{book2a,book2}.
Within this realm, the theme of nonlinear waves has received a significant
amount of interest due to the wide tunability and range of possibilities
available in the ever expanding number of experimental setups;
see e.g. for recent reviews~\cite{emergent,revnonlin,rab,djf}.
In parallel, significant developments in the field of (discrete,
as well as continuum) nonlinear optics~\cite{yuri,motirev} have
offered a complementary thrust especially in the study of
solitary waves and their dynamics and interactions in such dispersive
media.

One of the particularly intriguing dynamical structures that
was mathematically proposed~\cite{manakov,ablowitz} 
and experimentally realized~\cite{seg1,seg2,seg3} originally in 
optical systems, but which has garnered considerable recent 
interest in atomic BECs consists of the dark-bright (DB) solitary waves.
The wide range of 
theoretical~\cite{buschanglin,DDB,kanna,rajendran,val,berloff,VB,Alvarez,vaspra,vasnjp,fot,yan} 
and also of experimental~\cite{sengdb,peterprl,peter1,peter2,peterpra,peter3}
studies on this theme is, arguably, warranted by the special
nature of this structure. More specifically, while bright solitons cannot
exist ``on their own'' in the repulsive inter-atomic interaction setting
where DBs arise, the dark solitary waves play the role of an effective 
potential that enables the bound state trapping of the bright component.
The result of this symbiosis is a robust DB structure which has been
experimentally monitored, e.g., to oscillate with the customary harmonic
confinement~\cite{sengdb,peter1}, to be spontaneously produced by 
counterflow experiments~\cite{peterprl}, and 
to form bound states~\cite{peter2}.  Additionally, SU$(2)$-rotated
variants of the DB solitons have also been experimentally
observed assuming the form of beating dark-dark solitons~\cite{peterpra,peter3}.
It should also be noted that two-dimensional generalizations of the
DB solitary waves have also been explored in the context of the
so-called vortex-bright (VB) solitary waves~\cite{skryabin,kody}
and have been observed in experiments~\cite{andersonexp}. Pairs of such VB states have
also been recently considered~\cite{pola}.

Our aim in the present work is to connect this setup of the DB solitary
waves (and especially of the bound states of their pairs)
with another notion that has been extensively explored in 
atomic BECs, as well as in optics, namely that of the double 
well potential (DWP). The prototypical realization of the latter
in the context of BEC relies on
the combination of a parabolic trap
 with a periodic potential.
The latter
can be created in the form of an 
optical lattice, by the interference of laser
beams illuminating the condensate~\cite{Morsch}. The use of a DWP created as
a trap for BEC (together with self-repulsive nonlinearity) has revealed
a wealth of new phenomena originally predicted
theoretically, as e.g. in~\cite{smerzi,smerzi1,smerzi2},
and subsequently observed in recent experiments~\cite{markus1}. These include
the tunneling (Josephson) oscillations for small numbers of atoms; 
for higher atom numbers, more exotic states such
as ones involving macroscopic quantum self-trapping were experimentally
observed. Prior to this work, as well as afterwards motivated by its
findings, a wide range of theoretical studies investigated such DWP
settings, examining issues such as few-mode reduced descriptions of
the dynamics and
symmetry-breaking bifurcations \cite{kiv2,mahmud,bam,Bergeman_2mode,infeld2,todd,theo}, quantum effects \cite{carr,julia,bettina,julia2}, and
(purely) nonlinear variants of the DWP \cite{pseudo}, among others. 
DWP settings
and especially a signature feature therein, namely the potential
presence of spontaneous-symmetry-breaking effects  have also been
studied in nonlinear-optical settings. There, the formation of asymmetric
states in dual-core optical fibers \cite{fibers}, self-guided laser beams in
cubically nonlinear (so-called Kerr)
media \cite{HaeltermannPRL02}, and optically-induced dual-core waveguiding
structures in photorefractive crystals \cite{zhigang} have been
reported. In fact, it is this wide range of activities in the fields
predominantly of optical and atomic physics that have led to
books specifically dedicated to this subject~\cite{boris_book}.
Recently, further experimental activity has ensured a renewed interest
in this theme. In particular,  
the setting of two different hyperfine states of $^{87}$Rb 
was used to provide
a remarkable manifestation of the symmetry-breaking transition
(i.e., the pitchfork bifurcation arising in the double
well because of the effective nonlinearity imposed by interatomic
interactions)~\cite{zibold}. This enabled a full experimental mapping of
the 
phase plane of the double well system with un-paralleled accuracy. 
Additionally, extensions of the double well
setting were considered in more complex higher dimensional
settings such as a transversely elongated double well formulated
within an atom chip in~\cite{leblanc}.

The scope of the present work is to explore the connections between
the two above settings specifically by considering the realm
of two DB solitary waves and their existence, stability properties
and nonlinear dynamics. We find, following up on the earlier investigation
of~\cite{peter2}, that there are two possible stationary states consisting
of DB soliton pairs: one in which the bright ``pulses'' are in phase,
and one in which they are out of phase. However, the latter configuration
undergoes a symmetry breaking, destabilizing pitchfork bifurcation,
which, in turn, produces a third branch whereby the bright components
are asymmetric. This observation prompts us to examine the problem
in the realm of double well potentials and two-mode approximations:
more specifically, we consider the dark solitons as imposing 
(when combined with the parabolic trap) the effective DWP within
which the bright component finds itself. By reverting to this
effective DWP setting, we are able to make explicit analytical
predictions for the observed bifurcation that corroborate the full
numerical observations. More generally, we are able to reconstruct
the entire phase space in the plane of relative population imbalance versus relative phase of the two bright wavepackets. There is, however, one major difference compared to the static DWP which is the fact that in the present case the effective double-well is composed of a dynamic entity that is explicitly time dependent. Numerous investigations \cite{watanabe1}-\cite{watanabe2} have shown that an externally imprinted driving of the DWP greatly affects the dynamics, leading for instance to stimulation or complete destruction of tunneling \cite{watanabe1,strzys}, chaotic tunneling currents \cite{abdullaev,boukobza} or the creation of maximally entangled states on the many-body level \cite{watanabe2}. In our case, the situation is even more involved, since the dark solitons effectively forming the double well obey their own equation of motion that is affected by the dynamics of the bright component inside this double well.   
We explore parameter regimes where this feedback effect between the two components strongly modifies the dynamics, steering away from the effective one-component picture and
into a full two-component setting, where anharmonic and aperiodic motions
can be observed. We will see that this last set of features distinguishes
the DB case from its VB analogue where many of these features
(tunneling, symmetry-breaking, etc.) can still be observed~\cite{pola},
but the vortices are less responsive to the bright component's dynamics,
forming a comparably fixed background substrate.
Our presentation of the above features will be structured as follows.
In section II, we introduce the theoretical model of interest, while
in section III, we study the stationary states of the system and
the symmetry-breaking bifurcations. In section IV, we analyze
the system dynamics in the realm that can be captured by the two-mode,
effective one-component approximation. In section V, we explore the
limitations of the effective picture, by studying fully two-component
dynamical scenarios. Finally, in section VI, we summarize our findings
and present our conclusions.

\section{Theoretical Model}
Dark-Bright (DB) solitons emerge in binary repulsive Bose-Einstein condensates 
in the nK temperature regime. In that regime, they can be described by 
mean-field theory as particular solutions of the coupled two-component 
Gross-Pitaevskii equation which is given for a quasi one dimensional setup by 
\begin{align}
i\partial_{t} \psi_{D} &= \bigg[ -\frac{1}{2}\partial_{x}^2 +V_{\text{ext}}+ g_{D} |\psi_{D}|^2 + \sigma |\psi_{B}|^2 \bigg] \psi_{D} \notag \\
i\partial_{t} \psi_{B} &= \bigg[ -\frac{1}{2}\partial_{x}^2 +V_{\text{ext}}+ g_{B} |\psi_{B}|^2 + \sigma |\psi_{D}|^2 \bigg] \psi_{B} \label{ceq} 
\end{align}
where $\psi_{D,B}(x,t)$ are the condensate wavefunctions and the subscripts $D,B$ denote the dark and bright component respectively.

We choose a two-component mixture of $^{87}$Rb atoms in the two hyperfine states $F=1,\,m_{F}=-1$ and $F=2,\,m_{F}=1$ without interconversion between the two species, so that particle numbers in the two components are separately conserved. A reduction to a quasi-one-dimensional BEC leads for this choice to dimensionless coupling constants $g_{j}=a_{j}/|a_{DB}|=0.97,1.03$, where $j\in \{ D,B \}$ are the intra-component and $\sigma = 1$ the inter-component interaction strengths characterized by the corresponding $s$-wave scattering lengths $a_{D},a_{B},a_{DB}$.
$V_{\text{ext}}=\frac{1}{2}\alpha^2 x^2$ denotes the external harmonic trapping potential where $\alpha = \omega_{x}/\omega_{r}$ is the ratio of the longitudinal and radial trapping frequencies set to $0.2$ throughout this work. Time, length, energy and densities are measured in units of $\omega_{r}^{-1}$, $a_{r}$, $\hbar \omega_{r}$ and $(2|a_{DB}|)^{-1}$ respectively where $r=\sqrt{y^2+z^2}$ and $a_{r}=\sqrt{\hbar/m\omega_{r}}$ is the radial harmonic oscillator length~\cite{emergent}. In the following all results are presented in dimensionless units. 

Through the separation ansatz $\psi_{j}(x,t)=\exp{(-i\mu_{j}t)}\phi_{j}(x),\,\,j\in \{ D,B \}$ where $\mu_{D}$ and $\mu_{B}$ are the chemical potentials of the dark and bright component respectively, one obtains the binary stationary GPE.
A typical example of the density and phase profiles of the obtained
solutions is shown in Fig.~\ref{fig1}. 
Stationary DB solutions of Eq.~(\ref{ceq}) can be found through continuation from the linear limit through the use of an iterative Newton-Raphson solver~\cite{newtonsolver}. 
As trial wavefunctions we choose the lowest harmonic oscillator eigenfunctions where for the dark component we take the second excited state with two nodes and for the bright component either the ground state solution with no node that leads to the in-phase DB soliton or the first excited state with one node that leads to the out-of-phase DB soliton solution. To investigate the dynamics we employ a 4th order Runge-Kutta algorithm.

Furthermore, in order to test the robustness of the stationary solutions, 
we employ a linear stability or Bogoliubov de-Gennes (BdG) analysis which consists of slightly perturbing the stationary solutions through the ansatz
\begin{align}
\psi_{B}(x,t)=e^{-i\mu_{B}t}\bigg[  \phi_{B}(x)+\epsilon\sum_{k}\bigg( u_{k}(x)e^{-i\omega_{k}t}+v_{k}^{*}(x)e^{i\omega^{*}_{k}t} \bigg) \bigg] \notag \\
\psi_{D}(x,t)=e^{-i\mu_{D}t}\bigg[  \phi_{D}(x)+\epsilon\sum_{k}\bigg( a_{k}(x)e^{-i\omega_{k}t}+b_{k}^{*}(x)e^{i\omega^{*}_{k}t} \bigg) \bigg]
\end{align}
where $\phi_{B},\,\phi_{D}$ are stationary solutions. Inserting the above ansatz into Eq.(\ref{ceq}) and linearizing it with respect to the amplitude $\epsilon \ll 1$ of the excitations leads to an eigenvalue problem with $u_{k},\, v_{k},\, a_{k}, \, b_{k}$ constituting the eigenfunctions of the excitations to be determined and $\omega_{k}$ the (generally complex) eigenfrequencies. The appearance of eigenfrequencies $\omega$ 
with a non-vanishing imaginary part implies dynamical instability. 
The Hamiltonian structure of the underlying problem leads to quartets of
such eigenfrequencies (i.e., if $\omega$ is an eigenfrequency, so are 
$-\omega$, $\omega^{*}$ and $-\omega^{*}$), hence if Im$(\omega) \neq 0$,
there will always be a mode leading to growth and eventual deformation
of the configuration.

\section{Stationary DB Solitons and Symmetry Breaking Bifurcations}
\begin{figure}[!ht]
\centering
\includegraphics[width=0.8\textwidth]{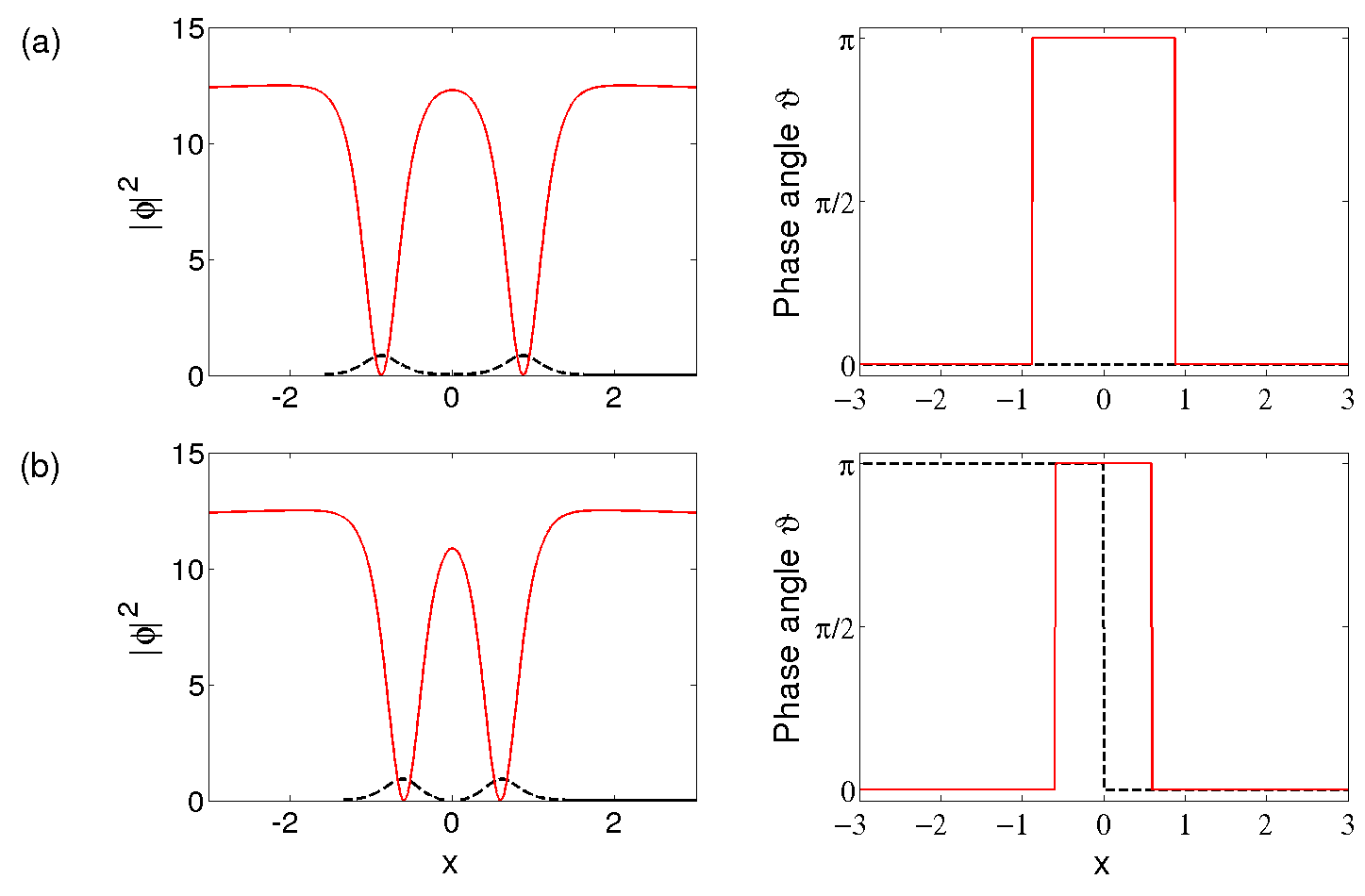}
\caption{Stationary DB soliton solutions of Eqs.~(\ref{ceq}). The red solid lines correspond to the dark component, the black dashed lines to the bright component. (a) density of in-phase DB solitons with $\Delta \vartheta = 0$ phase difference in the bright component (b) density of out-of-phase DB solitons with $\Delta \vartheta = \pi$ phase difference in the bright component, both figures with $N_{D}=400$ and $N_{B}=1$. On the right side, the corresponding phases are depicted.}
\label{fig1}
\end{figure}

There exist two different stationary DB solutions characterized by the phase difference $\Delta\vartheta$ between the bright solitons, the in-phase or symmetric solution with $\Delta \vartheta=0$ and the out-of-phase or antisymmetric solution with $\Delta \vartheta=\pi$~\cite{peter2}. As was shown in that work, 
and as is visible in Fig.~\ref{fig1}, the equilibrium distance between 
the two DB solitons crucially depends on the relative phase of the bright 
solitons, 
since their interaction is repulsive for $\Delta \vartheta=0$ and attractive for $\Delta \vartheta=\pi$ leading to a larger equilibrium distance in the in-phase solution. 
As will be seen below, this phase-sensitive DB soliton interaction is of fundamental significance in
the present work.

\begin{figure}[!ht]
\centering
\includegraphics[width=1.0\textwidth]{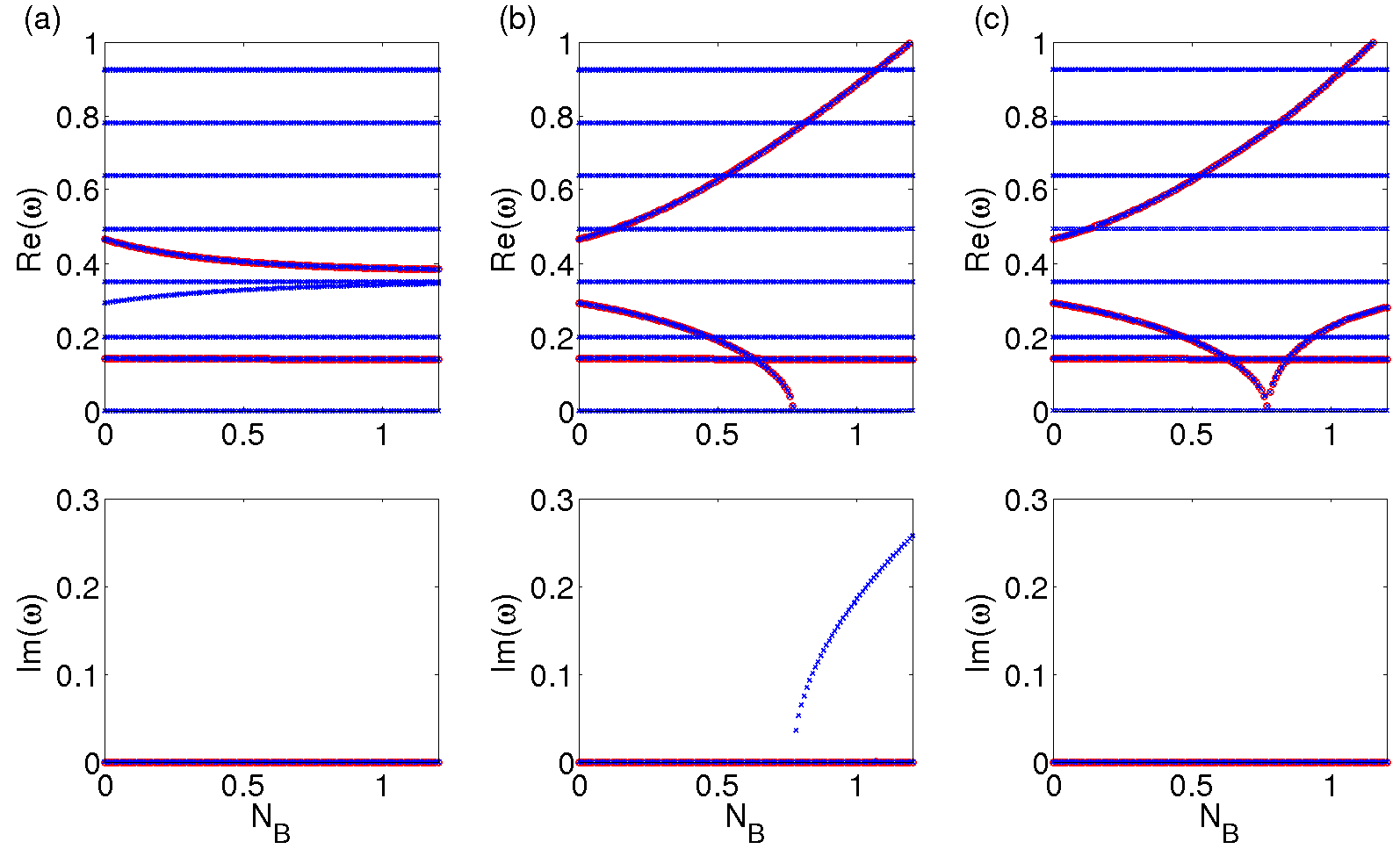}
\caption{BdG spectrum of the (a) symmetric, (b) anti-symmetric and 
(c) asymmetric stationary solutions of Eq~(\ref{ceq}) for $N_{D}=400$. 
Notice that the asymmetric branch only exists as such past the
critical point of the eigenfrequency zero-crossing. Below this value
of $N_B$, it coincides with the anti-symmetric branch.
The top panels
show the real part Re$(\omega)$ of the eigenfrequency $\omega$, 
while the bottom ones show the imaginary part Im$(\omega)$. When the
latter possesses non-vanishing values, an instability arises.}
\label{fig2}
\end{figure}

\begin{figure}[!ht]
\centering
\includegraphics[width=0.8\textwidth]{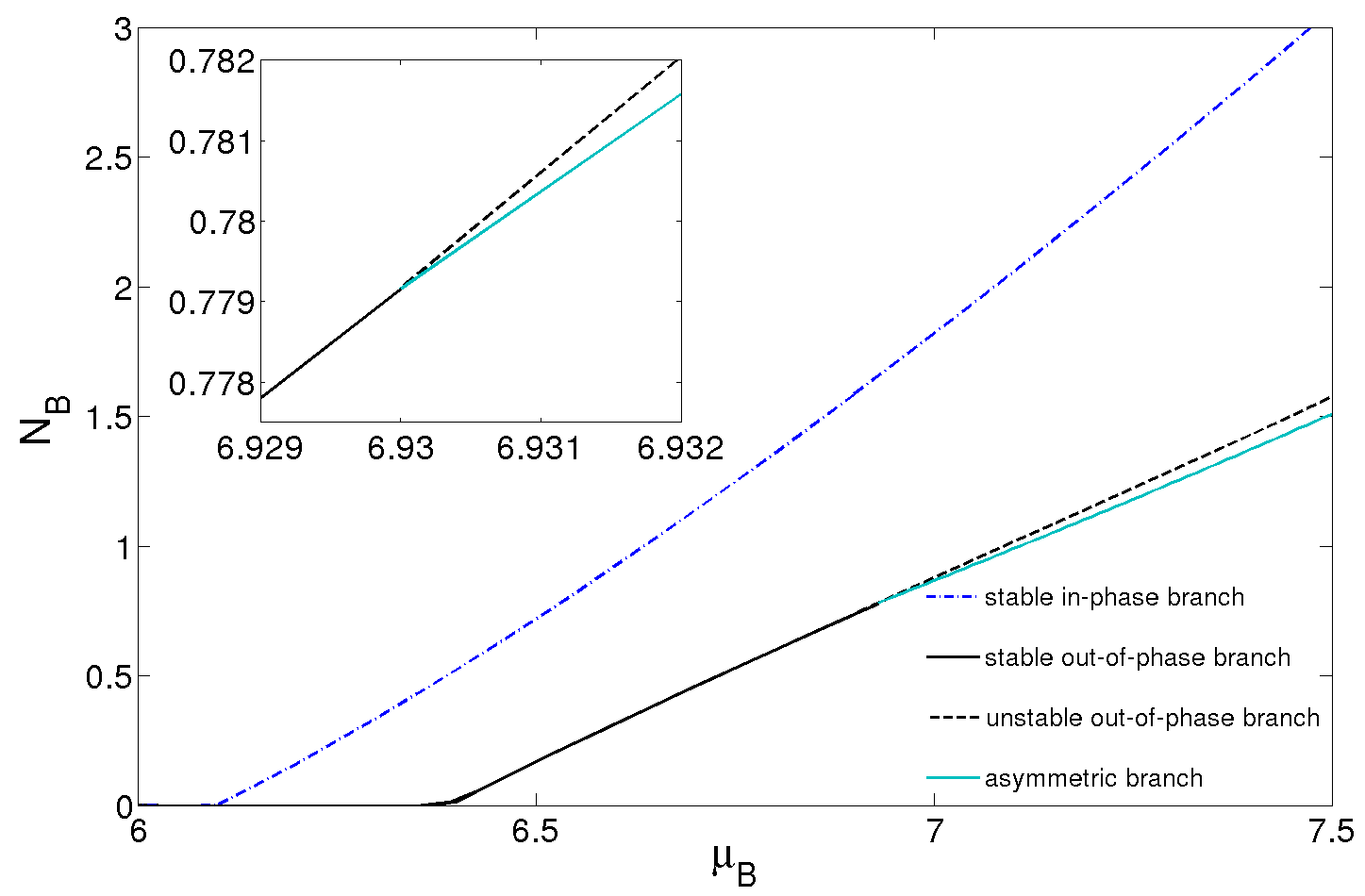}
\caption{Bifurcation diagram for $N_{D}=400$ and varying $N_{B}$ and $\mu_{B}$
for the stationary solutions of Eq.~(\ref{ceq}). The critical point where the pitchfork bifurcation occurs is $N_{\text{cr}}=0.779$ and $\mu_{\text{cr}}=6.93$. The inset shows a magnified view of the bifurcation point for better visibility.}
\label{fig3}
\end{figure}

The BdG analysis for the in-phase and out-of-phase solutions is shown in 
Fig.~\ref{fig2}(a) and (b), respectively. 
As has been discussed in detail in numerous works (see, e.g., the review
of~\cite{djf}), the DBs constitute excited states of the system, a feature
that is mirrored (in their linearization) through the emergence of so-called
negative energy, or negative Krein signature modes. For instance, for
dark solitons in one-component BECs, it was proved that an $n$-soliton
state carries $n$ such modes~\cite{chaostodd}. 
These modes are particularly important since their
potential collision with opposite (positive) Krein signature modes
gives rise to structural instabilities in the form of the so-called oscillatory
instabilities or Hamiltonian-Hopf bifurcations.
In the present setting the negative Krein signature modes, also called anomalous modes, are depicted by red color in the BdG analysis of Fig.~\ref{fig2}. 
Interestingly, the symmetric, in-phase solution in Fig.~\ref{fig2}(a) 
possesses two anomalous modes which correspond to particle like oscillations 
(in- or out-of-phase) of the DB solitons in the trap. 
This solution remains linearly stable in the considered regime
of parameters, as can be seen from the lower panel that shows the imaginary 
part of the frequency spectrum which is zero. In contrast to this, 
Fig.~\ref{fig2}(b), representing the spectrum of the out-of-phase 
solution, possesses three anomalous modes. 
The third anomalous mode that exists in this solution is due to the excited nature of the bright component wavefunction,
which in the overall linear limit reduces to the first excited harmonic oscillator state.
The dynamics of this third mode is found to correspond to an oscillating population imbalance between
the two bright wave packets. This was first observed in the dissipative
GPE framework (for DB pairs) of~\cite{vasnjp}, where the third mode
eigenfunction was added to the exact solution leading to the indicated
imbalance. 
The out-of-phase bright components attract each other and in this case a negative energy linearization mode 
emerges that captures the tunneling
of a fraction of the atoms from one atomic cloud (i.e., the one bright
solitary wave) to the other.
An important additional feature of this mode is that over parametric
variations (such as the number of bright atoms $N_B$), the mode
may cross the origin of the spectral plane as in Fig.~\ref{fig2}(b).
The resulting instability leads to a new, asymmetric stationary solution where the population of the bright component in the two dark solitons is different. 
Following the antisymmetric (out-of-phase) 
branch to the point where it gets unstable and perturbing it with the 
corresponding instability 
eigenvector leads therefore to a new branch of stationary solutions whose BdG spectrum is shown in Fig.~\ref{fig2}(c); this asymmetric solution
only exists past the relevant critical point of the eigenvalue 
zero-crossing while below it is identical to the antisymmetric branch. It possesses again three anomalous modes as the parental branch, but this new solution is linearly stable for all values of $N_{B}$ considered
as can be seen from the imaginary part of the 
spectrum. At the point where the instability sets in in Fig.~\ref{fig2}(b), a symmetry breaking bifurcation of the pitchfork type occurs and the out-of-phase (anti-symmetric) 
branch gets unstable whereas a pair of new, stable asymmetric branches 
emerges. 
This bifurcation is shown in more detail in Fig.~\ref{fig3} where all three branches, in-phase, out-of-phase and the new, asymmetric one are depicted. 
The symmetric branch shows no bifurcation as in the BdG spectrum whereas the antisymmetric branch bifurcates into the asymmetric one at the bifurcation point at $N_{\text{cr}}=0.779$ which coincides with the point where the imaginary mode emerges in Fig.~\ref{fig2}(b).

\begin{figure}[!ht]
\centering
\includegraphics[width=0.8\textwidth]{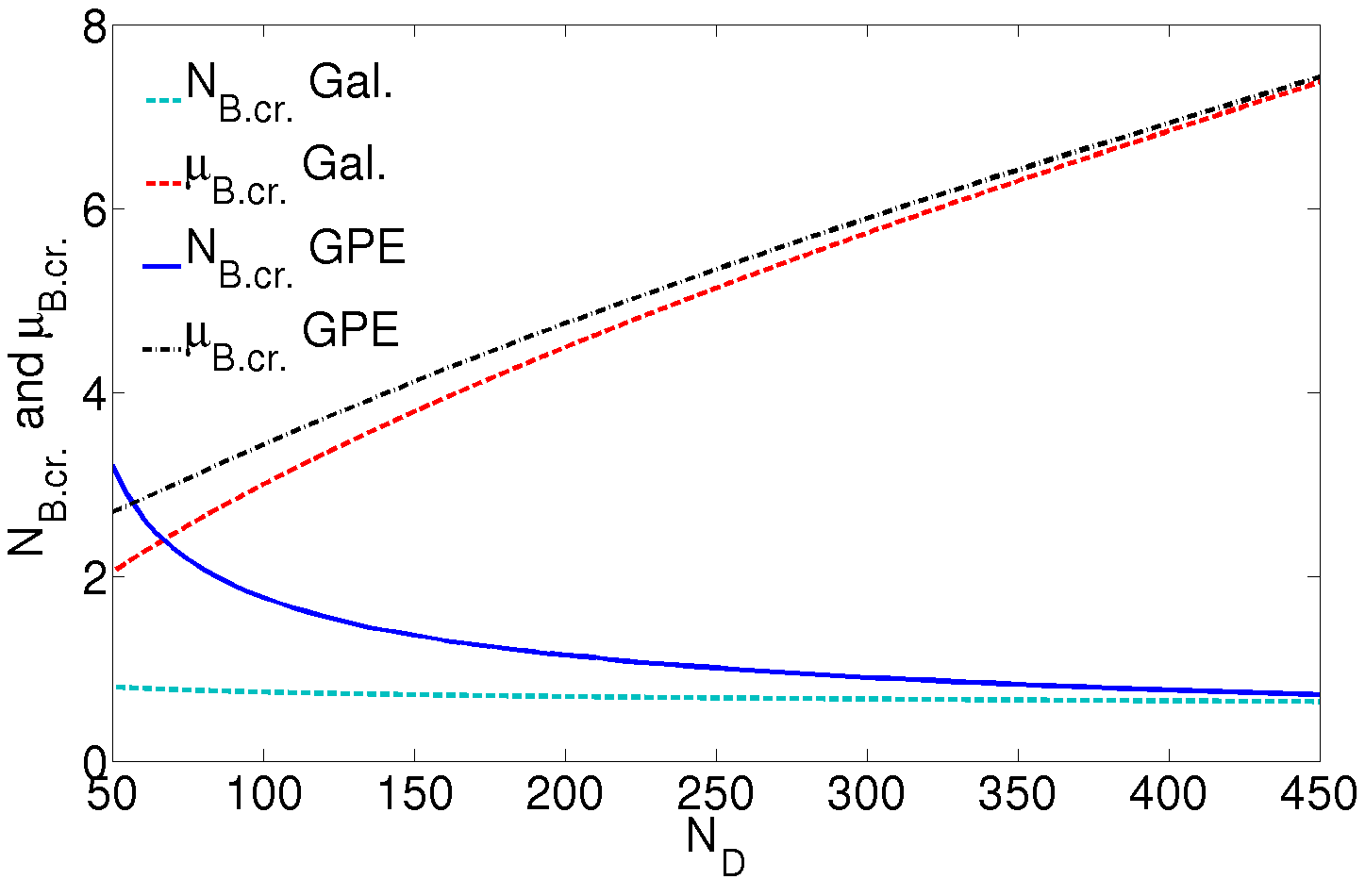}
\caption{Comparison of critical particle numbers and chemical potentials as predicted by full GPE calculations of Eq.~(\ref{ceq}) 
and the Galerkin two-mode approach for different $N_{D}$ as explained in the text.}
\label{fig4}
\end{figure}

This phenomenology is strongly reminiscent of the symmetry-breaking
bifurcation in a DWP and hence inspires a theoretical approach to understand the relevant features. 
More specifically, in the case of negligible feedback of the
bright component dynamics onto the dark component, another way to look at this symbiotic entity of the DB solitons
can be considered as follows. Assuming a given (stationary or time-dependent) dark component wavefunction $\psi_{D}(x,t)$ one may write $V_{\text{eff}}(x,t)=V_{\text{ext}}(x)+|\psi_{D}(x,t)|^2$ and consider only the bright species in a single component problem with the effective GPE
\begin{equation}
i\partial_{t}\psi_{B}=\Big[ -\frac{1}{2}\partial_{x}^2+V_{\text{eff}}+g_{B}|\psi_{B}|^2 \Big]\psi_{B} \label{ceqb}
\end{equation}
As long as the time scales on which the dark solitons move and the tunneling dynamics of the bright component takes place are well separated, the effective double well potential in the equation above can be approximated as being time independent.
In other parameter regimes, the time dependence of the effective potential term and the dynamic feedback on $\psi_D$ cannot be anymore neglected and the equation of motion of the dark solitons has to be included, that is the full two component GPE in Eqs.~(\ref{ceq}) has to be taken into account.
Let us focus for now on the first case where feedback is negligible and the solitons are essentially at rest (such that the effective potential can be taken as time independent).
We aim at a theoretical understanding of the bifurcation diagram shown in Fig.~\ref{fig3} making use of the effective one-component picture for $\psi_B$. As can been seen in Fig.~\ref{fig1} already, the
dark component wavefunction differs substantially for the in-phase and out-of-phase branch of the bifurcation diagram, due to the qualitatively different interaction between the DB soliton entities.
Thus, for each of the two branches one has to use the appropriate $\psi_D$ in $V_\text{eff}$, taking into account the correct dark soliton equilibrium distance. 
Then, for each of the two branches, $V_\text{eff}$ is given by an effective DWP formed by the two solitons plus the harmonic trap,
and solving Eq.~(\ref{ceqb}) yields two lowest eigenstates with even and odd parity, separated from all the higher eigenstates by a suitable energy bandgap. The idea then, in line with
similar explorations in the DWP realm~\cite{smerzi,Bergeman_2mode,todd,theo}
is therefore to use a two-mode ansatz to describe the full wavefunctions 
near the linear limit given by $\psi_{B}(x,t)=c_{0}(t)\phi_{0}(x)+c_{1}(t)\phi_{1}(x)$
where $\phi_{0}(x),\phi_{1}(x)$ are the 
normalized symmetric and anti-symmetric eigenfunctions of the lowest 
two energy levels in the effective double well, respectively, 
and to use again an iterative continuation to reach larger 
values of $N_B$.
Inserting this ansatz into Eq.~(\ref{ceqb}), one can 
then use the analysis of the two mode expansion developed earlier
(cf. e.g.~\cite{todd,theo}) to deduce the critical values in the bright 
component
particle number for which a symmetry breaking bifurcation of the pitchfork 
type emerges. In the symmetric branch no such bifurcation occurs 
whereas the anti-symmetric one becomes unstable at~\cite{theo}
\begin{eqnarray}
N_{\text{cr}} = \frac{\Delta \omega}{g_{B}(3A_{0011}-A_{1111})}; \quad
A_{0011} = \int_{-\infty}^{\infty} \mathrm{d} x \, \phi_{0}^2\phi_{1}^2; \quad 
A_{1111} = \int_{-\infty}^{\infty} \mathrm{d} x \, \phi_{1}^4
\label{ncr}
\end{eqnarray}
with 
the critical value of the chemical potential
$\mu_{\text{cr}} = \omega_{0}+3A_{0011}N_{\text{cr}}$; $\Delta \omega$ is the energy difference between the lowest two eigenvalues of $V_{\text{eff}}$ and 
the
 overlap integrals used in Eq.~(\ref{ncr}) utilize the 
symmetric and anti-symmetric eigenfunctions of the effective
potential $V_{\text{eff}}$. 
In contrast to the analysis usually carried out for static double-well potentials, where both the symmetric and
the anti-symmetric solution exist in one and the same external potential, 
a particular feature of the present setting is that 
we have to work with different double-well
potential shapes for the two branches, due to the different dark soliton distances. 
We now apply Eq.~(\ref{ncr}) to estimate the critical values where the asymmetric branch emerges.
As an input for the two-mode approach, we numerically solve the full GPE (\ref{ceq}) for the out-of-phase DB soliton pair and extract the wavefunction of the dark component.
This is done for varying values of $N_D$, while we fix $N_B = 1$, anticipating the rough position of the bifurcation point (we have checked that a different choice of $N_B$ only weakly affects the results).
For each $N_D$ we then construct the effective potential, find its lowest eigenstates and obtain the critical values from Eq.~(\ref{ncr}).
A comparison of the bifurcation point as a function of $N_D$ found from the full GPE and predicted by this effective potential two-mode approach is shown in Fig.~\ref{fig4}; recall
that in the case of the in-phase solution, no new branch bifurcates. The agreement is more and more exact with increasing $N_{D}$. This can
be explained by the fact that for larger particle numbers in the dark component the effective double-well gets deeper
and the band gap becomes larger so that the two-mode approach in turn becomes more accurate.
As will be seen
in the dynamical simulations, the two-mode approaches based on these two different effective potentials are not only
capable of capturing their respective fixed points as shown in the bifurcation diagram, but also the BJJ phase space
in the vicinity of these fixed points (and only there).

\begin{figure}[!ht]
\centering
\includegraphics[width=0.8\textwidth]{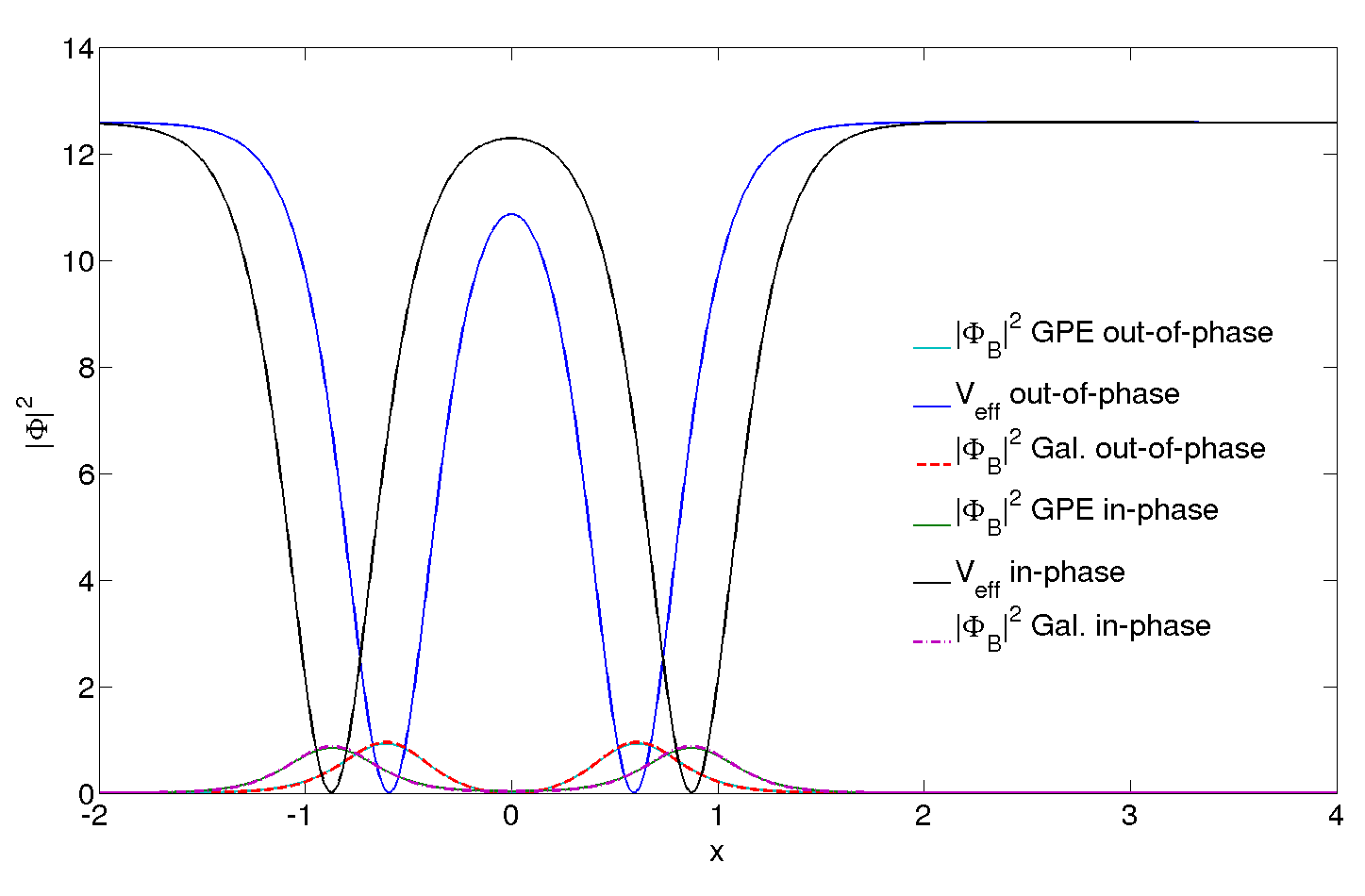}
\caption{Effective potentials for zero and $\pi$ phase difference in the bright component with their corresponding stationary solutions as calculated from the 
GPE and within the two-mode approach for $N_{D}=400$ and $N_{B}=1$. Despite the different distances of
the effective double-well potential, the appropriate two-mode approximation yields
a good approximation to the GPE results in both cases.}
\label{fig5}
\end{figure}

Within the two-mode picture one can start from the symmetric and anti-symmetric wavefunctions in the respective effective potential and construct wavefunctions that are localized in the right and left well through
\begin{equation}
\phi_{l,r}=\frac{\phi_{0}\pm \phi_{1}}{\sqrt{2}}
\end{equation}
With these localized wavefunctions one can expand the bright component in this basis according to
\begin{equation}
\psi_{B}=\sqrt{N_{l}}e^{i\theta_{l}}\phi_{l}+
\sqrt{N_{r}}e^{i\theta_{r}}\phi_{r} \label{two}
\end{equation}
where $N_{l,r}$ are the particle numbers in the left and right well respectively, $\theta_{l,r}$ the corresponding phases of the condensates in each well and $N_{B}=N_{l}+N_{r}$.\\
In Fig.~\ref{fig5} a typical example of the 
comparison of the stationary bright soliton solutions 
as constructed from the two-mode approach and from the full GPE simulation 
is shown. 
Both stationary wavefunctions, in-phase and out-of-phase, are very well approximated through the two-mode approach and importantly, the modes constructed
using the effective potential of the in-phase branch are not suitable for the out-of-phase branch and vice-versa (due
to the different displacement of the dark solitons in the two different
cases). This will lead in the following section to the result that we can observe different types of tunneling oscillations in the different effective potentials.

\section{Tunneling Dynamics}
\begin{figure}[!ht]
\centering
\includegraphics[width=1.0\textwidth]{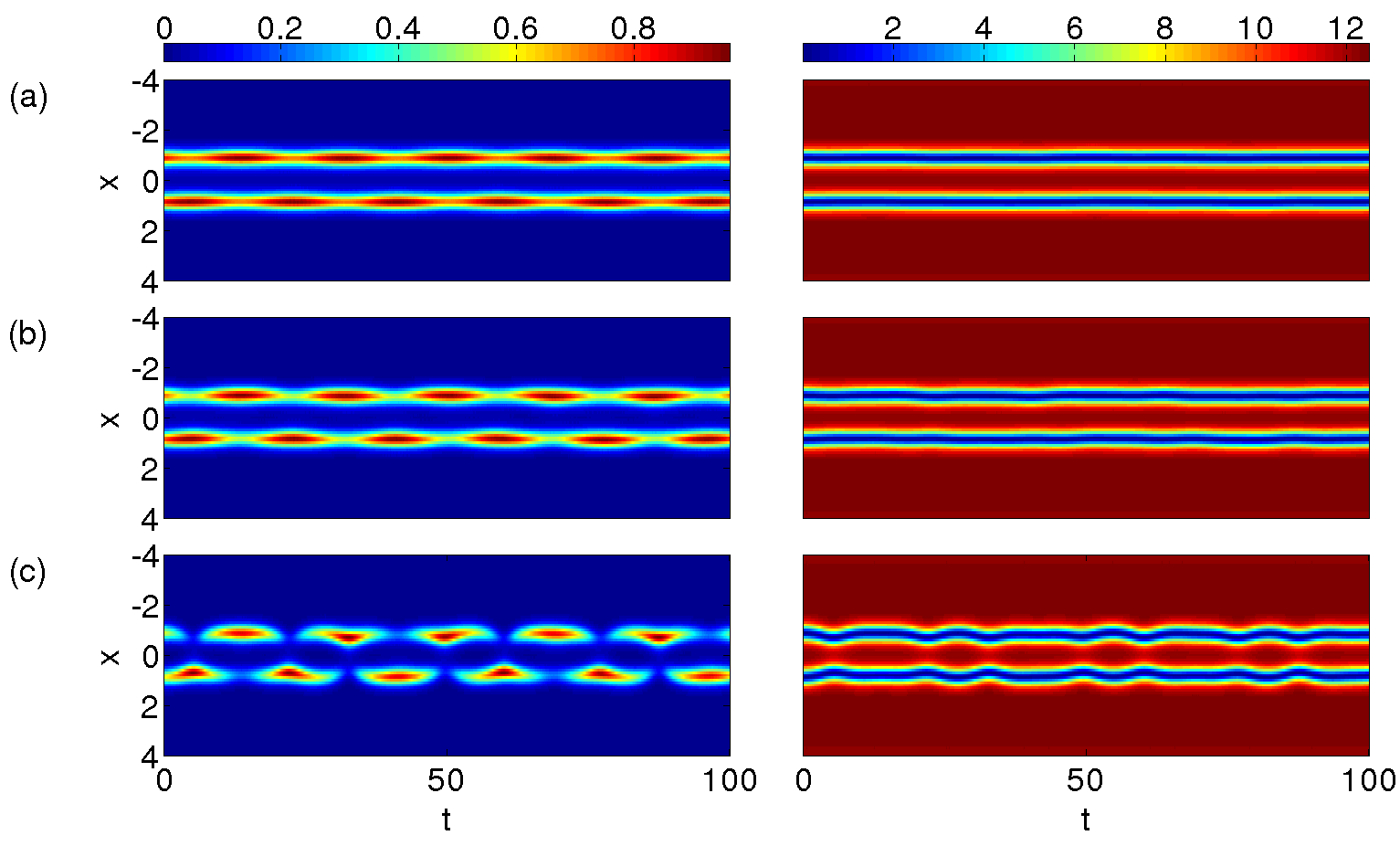}
\caption{Plasma oscillations around a phase difference of $\theta = 0$ with $N_{D}=400$, $N_{B}=1$ and initial conditions $z(0)=0$ and (a)
$\theta(0)=0.4$, (b) $\theta(0)=0.8$, (c) $\theta(0)=2.0$. Left panels show the density of the bright component, right panels the density of the dark component.}
\label{fig6}
\end{figure}

In the following we will focus on the case $N_{D}=400$ and $N_{B}=1$ where as one can see from Figs.~\ref{fig4} and \ref{fig5} the two-mode approach and the full GPE computation yield essentially identical results, as regards the 
steady state and stability analysis. In order to investigate the dynamics and in particular the feedback effect that arises through the coupling of the effective potentials with the bright component, we prepare the latter according to Eq.~(\ref{two}) and solve then Eqs.~(\ref{ceq}) where the time evolution of the effective potential is included.
Within the two-mode approach the reduced phase space of interest is $(\theta,z)$ with $\theta=\theta_{l}-\theta_{r}$ and $z=\frac{N_{l}-N_{r}}{N_{l}+N_{r}}$ being the phase difference and the population imbalance between the two wells;
see e.g.~\cite{smerzi,zibold}. For a static double well potential the particle numbers in the left and right well can be simply obtained by projecting onto the left and right localized wavefunctions in Eq.~(\ref{two}). Since in the full dynamical simulations this two-mode projection is not always applicable (e.g. since the dark solitons may start to move), we calculate the particle numbers through $N_{l}=\int_{-\infty}^{0} \mathrm{d}x \,  |\psi_{B}|^2 $ and $N_{r}=\int_{0}^{\infty} \mathrm{d}x \,  |\psi_{B}|^2 $ and the phases correspondingly through $\theta_{l}=\text{arg}(\int_{-\infty}^{0} \mathrm{d}x \,  \psi_{B})$ and $\theta_{r}=\text{arg}(\int_{0}^{\infty} \mathrm{d}x \,  \psi_{B})$. In the regime where the time dependence of the effective potentials is negligible both methods of obtaining particle numbers and phases yield equivalent results.
The two different effective potentials 
that arise from the in-phase and out-of-phase DB solitons support different 
tunneling oscillations. The in-phase effective potential supports a 
stationary solution with $\theta = 0$ which is a stable fixed point within the reduced phase space picture.  
The oscillations around this point are usually referred to
as plasma oscillations. 
Fig.~\ref{fig6} shows the dynamics of the densities of the dark and bright solitons for three different initial conditions of the phase 
difference and population imbalance. One sees that as the values 
(in this case of the relative phase) 
deviate from the stationary solution, the tunneling increases.
The bottom row is, arguably, the most important among the three panels;
it illustrates clearly the point discussed previously about the dark
solitons constituting a ``flexible substrate'' that eventually
(i.e., upon sufficient deviation from the equilibrium limit) responds
to the dynamics of the bright structures. As a result, the dark
solitons themselves start oscillating within the trap varying their
relative distance over time in a way coupled to the ongoing tunneling
dynamics.

\begin{figure}[!ht]
\centering
\includegraphics[width=1.0\textwidth]{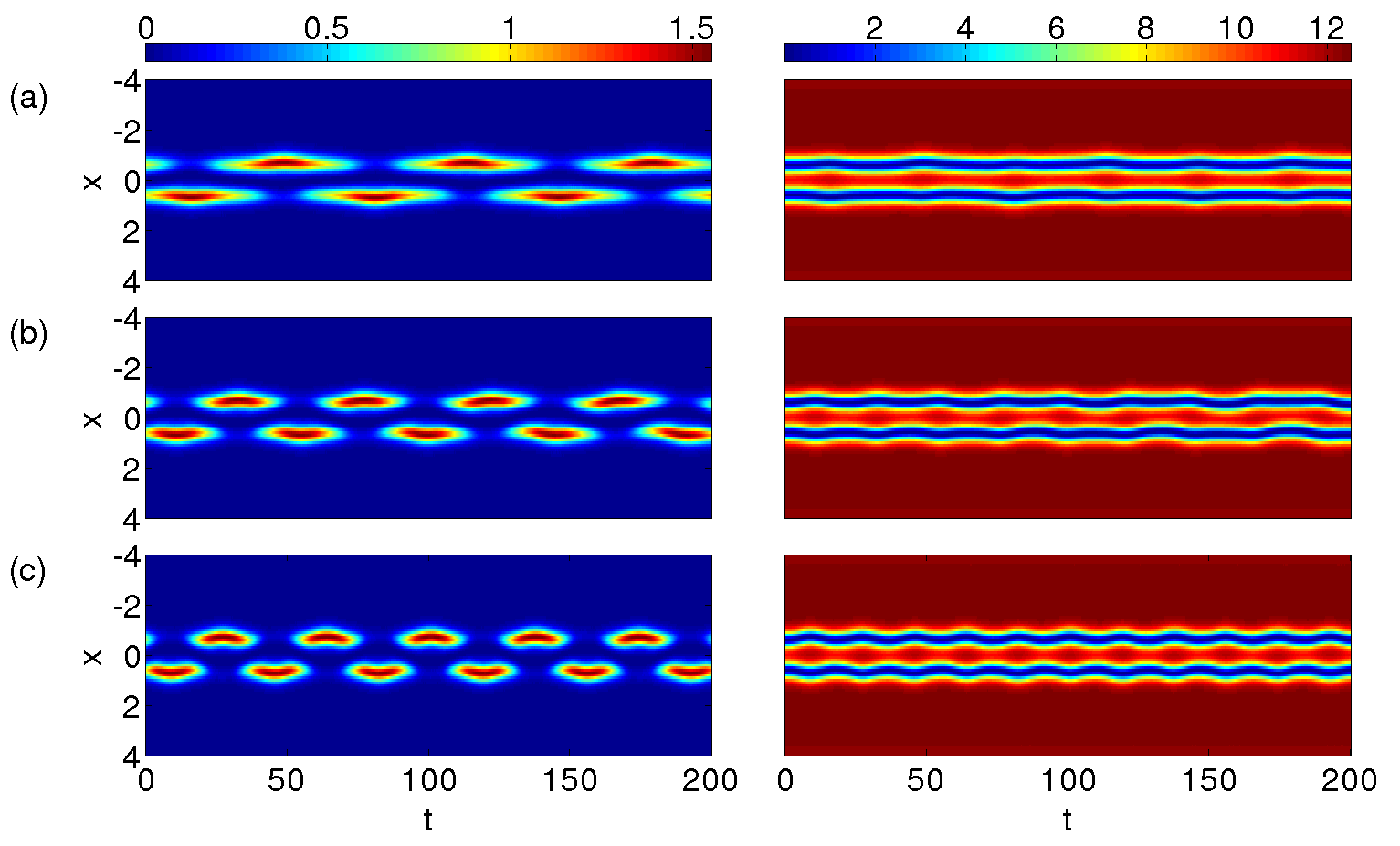}
\caption{$\pi$-oscillations around a mean phase $\theta =\pi$ with $N_{D}=400$, $N_{B}=1$ and initial conditions $z(0)=0$ and (a) $\theta(0)=3.2$, (b) $\theta(0)=3.3$, (c) $\theta(0)=3.4$. 
Left panels show the density of the bright component, right panels the density of the dark component.}
\label{fig7}
\end{figure}

On the other hand, the effective potential built from the out-of-phase 
stationary solution supports what is known as $\pi$-oscillations 
(see e.g.~\cite{smerzi,zibold}) 
with a phase difference $\theta = \pi$ between the bright solitons. 
These are illustrated in Fig.~\ref{fig7}, where again similar characteristics
are evident. Namely, closer to the fixed point, the feedback effect 
and the resulting oscillation of the dark solitons is nearly imperceptible
(although arguably stronger than the corresponding effect in the in-phase
case, presumably due to the structural instability of the relevant
fixed point). However, the effect becomes progressively more discernible
as the deviation from the relative phase of $\pi$ increases between
the two effective wells.

\begin{figure}[!ht]
\centering
\includegraphics[width=1.0\textwidth]{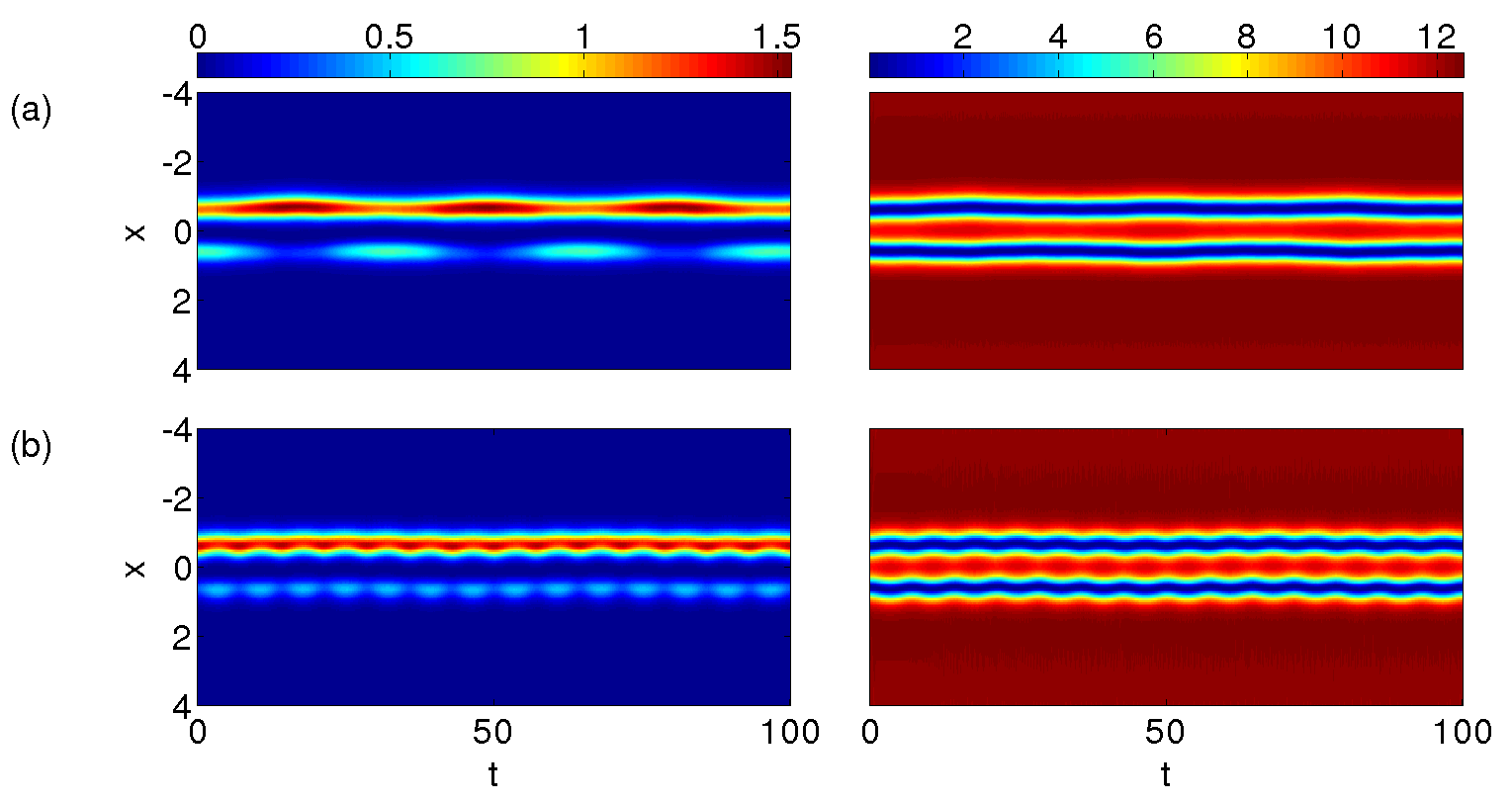}
\caption{Self-trapped oscillations with $N_{D}=400$, $N_{B}=1$ and initial conditions $\theta(0)=\pi$ and (a) $z(0)=0.3$, (b) $z(0)=0.7$. Left panels show the density of the bright component, right panels the density of the dark component.}
\label{fig8}
\end{figure}

The third and last relevant type of oscillations consists of the 
so-called self-trapped oscillations~\cite{smerzi,zibold}. 
They occur after initializing the system close to one of the new, stable asymmetric fixed points with $z \neq 0,\, \theta=\pi$ (i.e., at a non-vanishing population imbalance).
This is, in turn, shown in Fig.~\ref{fig8}, in this case for fixed
relative phase but for different population asymmetries. Here too,
the departure from the relevant fixed point forces the dark solitons
performing the trapping to oscillate together with the bright atomic
wavepackets they are trapping in an out-of-phase fashion.

\begin{figure}[!ht]
\centering
\includegraphics[width=1.0\textwidth]{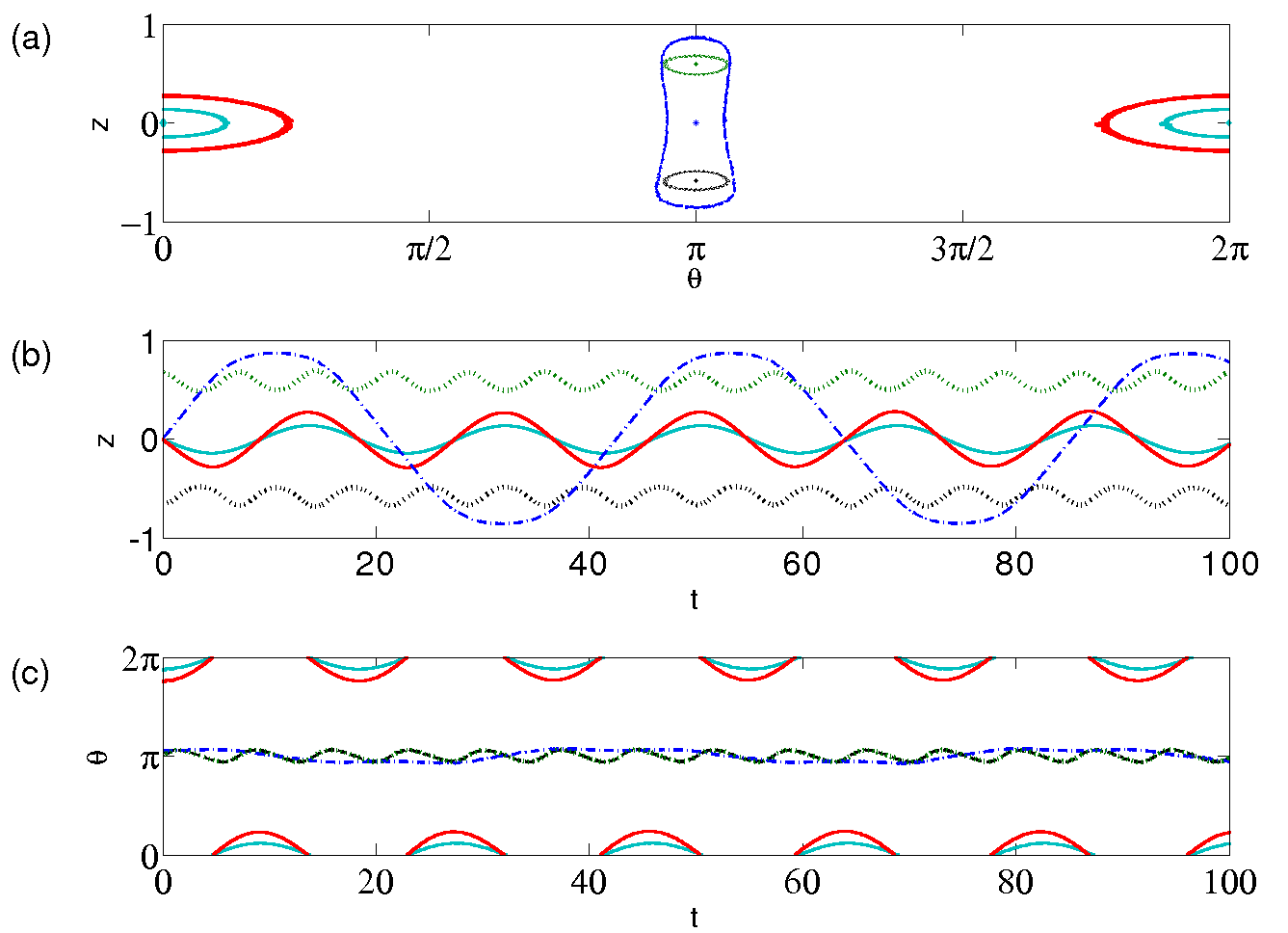}
\caption{(a) Phase space of the effective double well potential, as 
reconstructed from the dynamics of Eq.~(\ref{ceq}) for 
$N_{D}=400$ and $N_{B}=1$, (b) Time evolution of population imbalance $z$, 
(c) Time evolution of phase difference $\theta$. The red and 
cyan curves correspond to plasma oscillations shown in Fig.~\ref{fig6}(a) and (b), the blue to $\pi$-oscillations shown in 
Fig.~\ref{fig7}(b) and the black and green to self-trapped oscillations 
shown in Fig.~\ref{fig8}(b).}
\label{fig9}
\end{figure}

We collect the phase space trajectories corresponding to these different types of tunneling oscillations in Fig.~\ref{fig9}. 
It resembles the characteristic features of the BJJ phase space, that has been previously
constructed in one-component systems theoretically~\cite{smerzi} and
experimentally~\cite{zibold}.  
Yet, it has to be noted that the two-mode projections underlying the phase space trajectories around the symmetric fixed point and the anti-symmetric/asymmetric
fixed points are in fact different here, one being based on the in-phase effective DWP, the other one on the out-of-phase
effective DWP. We remark that temporal evolution of the population imbalance 
shown in the middle panel can be accurately described by
means of harmonic or elliptic oscillations as in the case of a static double-well potential~\cite{smerzi}. In the lower panel, the 
phase oscillations associated with the dynamical evolution 
are depicted. Corroborating the earlier picture, these oscillate  
around a mean phase difference of $\theta = 0$ for the plasma oscillations, 
or around $\theta = \pi$ for $\pi$- and self trapped oscillations. 

\section{Limitations of the Two-Mode Approach}
\begin{figure}[!ht]
\centering
\includegraphics[width=0.6\textwidth]{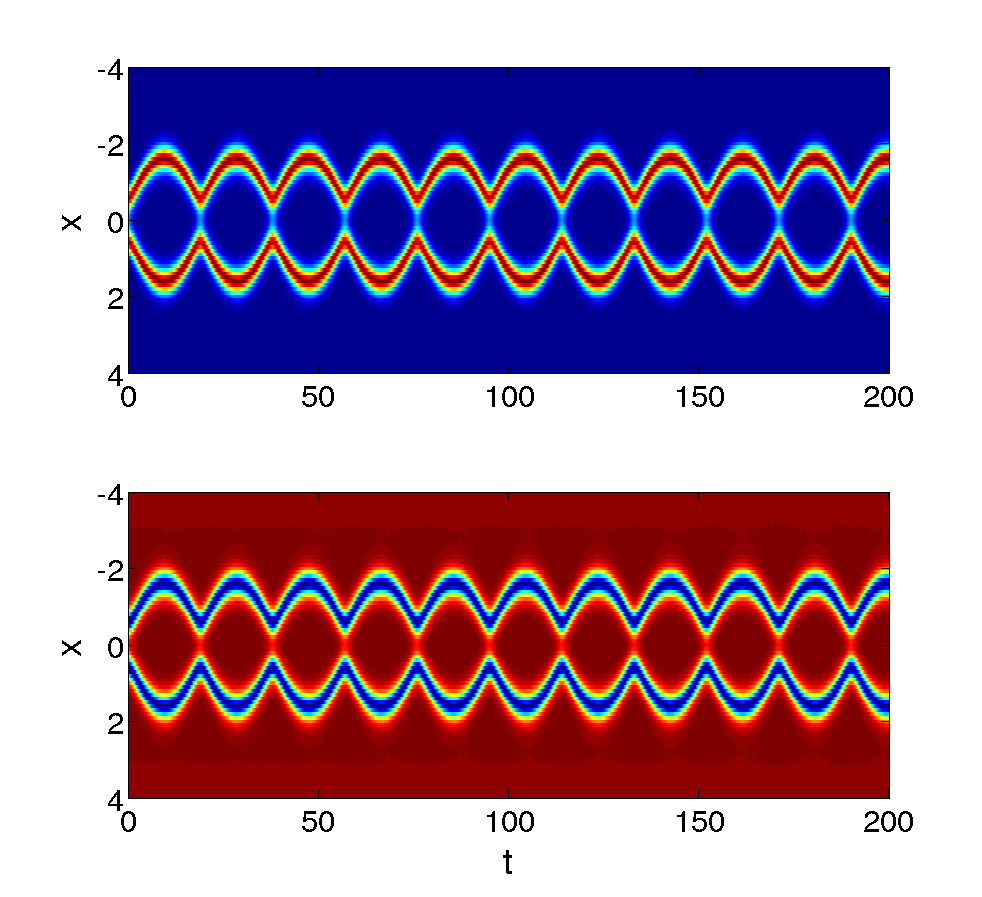}
\caption{Dynamics following a phase quench in the bright component,
starting from the stationary out-of-phase solution and quenching the
phase to $\theta = 0$. Upper panel shows the density of the bright component, the lower panel that of the dark component.}
\label{fig10}
\end{figure}

\begin{figure}[!ht]
\centering
\includegraphics[width=1.0\textwidth]{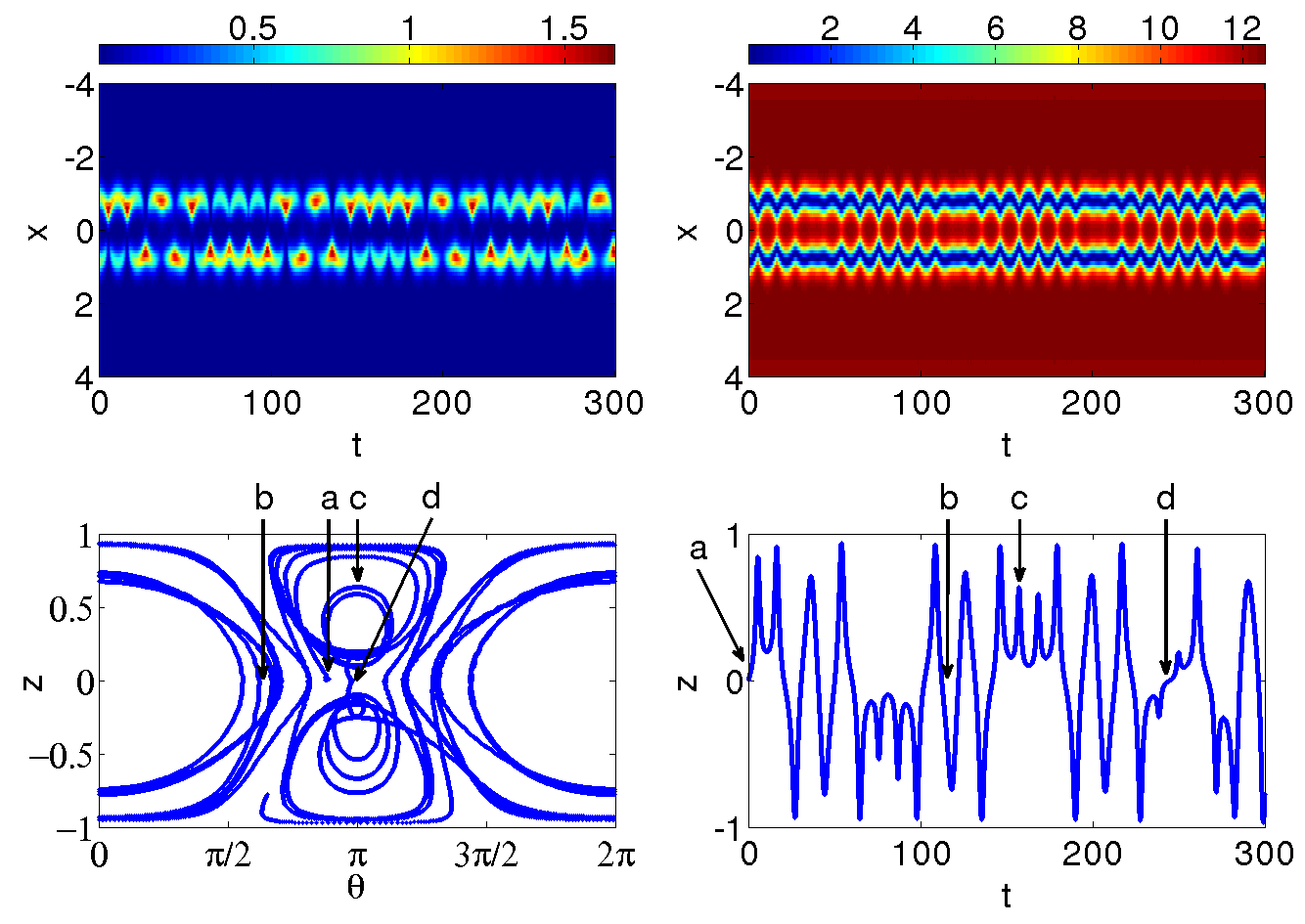}
\caption{ Dynamics following a phase quench in the bright component,
starting from the stationary in-phase solution and quenching the phase
to $\theta = 2.8$. The upper left panel shows the density of the bright
component, the upper right panel that of the dark component. The lower
left panel gives the reduced phase space picture and the lower right panel shows the population imbalance over time. Markers show the time-order in which the trajectory is traversed. The starting point corresponding to 'a' is the initial condition, 'b' is reminiscent of plasma oscillations, 'c' of self-trapped oscillations and 'd' of $\pi$-oscillations.}
\label{fig11}
\end{figure}

As demonstrated in the last section there exists a regime of initial conditions where all characteristic features of the phase space of the one-component DWP can be recovered. 
This illustrates the validity and the value of the effective one-component
picture (or, rather, of the two separate effective one-component pictures for the two branches) that we have put forth herein.
Nevertheless, as already illustrated in the bottom panels
of Figs.~\ref{fig6}--\ref{fig8}, there is also the possibility of 
choosing initial values lying far off the stationary states. 
The larger the deviation from the fixed points is, the less accurate
the effective one component reduction becomes. 
The DB soliton interaction crucially depends on the relative phase of the bright solitons. This means that continously through the tunneling process the phase difference and therefore the interaction strength change. This also 
implies the possibility that the nature of the interaction can change from  repulsive to attractive when a trajectory passes $\theta=\pi/2$ or $\theta=3\pi/2$. 
The initial effective
potentials underlying our two-mode approximations are built from stationary solutions with phase differences of $0$ or $\pi$ in the bright component. The increasing departure from these well-established
limits
can lead to such large amplitude oscillations that the reduction to
a two dimensional 
phase space is no longer adequate to describe the dynamics of the 
full infinite-dimensional problem.
Hence, while this simplified effective one-degree of freedom is particularly
insightful when applicable, it is natural to expect that the full 
infinite-dimensional dynamical system can bear a considerable additional
wealth of relevant dynamics. 
Some typical examples of this form are shown in 
Figs.~\ref{fig10} and \ref{fig11}. Here, the coupled nature of the
dark and bright solitary wave dynamical evolution is particularly
evident, leading the full system to periodic or aperiodic, apparently
``irregular'' dynamics. 
In Fig.~\ref{fig10}, we start from the stationary out-of-phase DB soliton pair solution and suddenly switch the phase difference in the bright component to $\theta=0$.
 This change in the phase difference leads to the above described qualitative change in the interaction from attractive to repulsive, so that the DB solitons abruptly feel a strong outward directed force. They start therefore to oscillate away from the trap center until the harmonic trap counterbalances this motion. 
During the dynamics, the parity symmetry of the bright component is fully preserved and there is no tunneling. 
The dynamics shown in Fig.~\ref{fig11} is initiated by starting from the in-phase stationary DB pair solution and suddenly switching the phase difference to a value close to $\pi$. The DB solitons experience a strong attractive force pointing towards the trap center and first start approaching each other. Tunneling oscillations of the bright component are initiated. Despite the complex dynamics seen in the full simulation of the coupled GPE system, within the two-mode projection one can still recognize parts of the old phase space shown in Fig.~\ref{fig9}(a). 
To illustrate this, markers have been added in Fig.~\ref{fig11} which assign some specific points in phase space and show the time-order in which the trajectory is passed.
Naturally, an insistence to describe these types of 
evolution via a reduced phase-plane picture leads to meaningless
results such as intersecting phase plane
curves in Fig.~\ref{fig11} clearly pointing out the insufficiency
of the reduced description to capture the GPE picture. 

\section{Conclusions \& Future Challenges}

In the present work, we have illustrated the generic existence
of two different stationary dark-bright soliton 
solutions in binary, repulsive, quasi-one-dimensional 
Bose-Einstein condensates.
One of these two states possesses two bright solitary waves
that are in-phase, while the other features an out-of-phase structure
between these wavepackets. An instability of the out-of-phase
two dark-bright soliton state was identified in the full system,
leading to a symmetry-breaking bifurcation of an asymmetric state.
Mathematically, this was
illustrated through the existence of an additional anomalous
(or negative Krein signature) mode associated with this asymmetry,
which beyond a critical nonlinearity crosses to the imaginary 
eigenfrequency axis.
To quantitatively address this symmetry-breaking pitchfork bifurcation,
we adopted an effective one-component picture wherein the dark solitons
were assumed as constituting a fixed double well potential, within
which the bright components evolved. This picture (when appropriately taking into account the dark soliton equilibrium distances) is very
accurate in the vicinity of the stationary states and enables
the observation of all the dynamical characteristics of the 
one-component double well potential system, including plasma
and $\pi$-oscillations, as well as (asymmetric) self-trapped states. 
As expected, this static double well framework cannot cover the full dynamics of the two-component GPE system,
and indeed it was found to break down far from the stationary states. From a physical
perspective, this breakdown arises due to the flexible nature
of the double-well substrate, namely the fact that the feedback
of the bright into the dark solitons leads the latter to dynamically
evolve and oscillate as well, rather than retain the character
as a fixed background potential.
From the perspective of dynamical
observations,  this phenomenon led to the manifestation of
other possible outcomes such as the periodic oscillation of the
entire dark-bright soliton pair, or the aperiodic motion thereof.
It is relevant to interject here a comment about the comparison of
these features to the ones observed in a quasi-two-dimensional analogue
of the system in~\cite{pola}.
In the analysis of two vortex-bright states in two dimensions 
the dark component's vortices were found to be less responsive to the 
bright component's tunneling dynamics, and if a slight particle-like 
precessional motion of the vortices was induced, its time scales 
were observed to be well separated from those of the tunneling oscillations.
This is no longer true for the dark-bright solitons in one dimension where regardless of the ratio 
$N_{D}/N_{B}$ the solitons can be found to move (under suitable
initial conditions). 
A fundamental difference between the two-dimensional and the one-dimensional case
is that the particle-type interaction between dark solitons is directed along their line of sight
(and thus aligned with the mass transfer of the tunneling bright component filling them), 
while the interaction between two vortices acts perpendicularly to their line of sight.
Furthermore, the topological nature of the vortices may impede a modification of their structural characteristics,
while this is certainly not impossible in the dark soliton case, where the
continuum of dark/gray soliton solutions may facilitate a back-action on the solitons forming the effective potential for the tunneling.
The present work opens several possibilities that
are worthwhile to consider for future studies. Although all of the
dynamics reported here are essentially localized in nature, we have
not attempted to utilize the dynamical picture put forth e.g.
in~\cite{peter2} about dark-bright soliton interactions in order
to attempt to understand the observed phenomenology. This is
certainly something that would be relevant to do, both at the
level of stationary states, but also at that of the stability
and of the observed dynamics. 
While the effective two-dimensional phase space reduction 
can be seen to be inadequate in certain regimes
herein, it would be natural to expect that still a simplifying description with
only few degrees of freedom (taking into account, e.g., also the dark soliton positions as dynamical variables)
may help in capturing the different features exhibited by the infinite dimensional GPE system.
A relevant effort would certainly be worthwhile
to undertake also in higher dimensions, including vortex-bright
solitary waves in 2d and vortex-ring-bright solitary waves
in 3d. In the case of the higher dimensional structures,
a more detailed understanding of the possible dynamical features
and of the interactions between the coherent structures is still
far from complete, although important recent steps are being
made in that direction~\cite{tsubota}. 

\begin{acknowledgments}
J.S. acknowledges support from the {\it Studienstiftung des deutschen
Volkes}. P.G.K. gratefully acknowledges the support of 
NSF-DMS-1312856, as well as from
the US-AFOSR under grant FA950-12-1-0332, the
Binational Science Foundation under grant 2010239, from the
Alexander von Humboldt Foundation and the ERC under FP7, Marie
Curie Actions, People, International Research Staff
Exchange Scheme (IRSES-606096).
PGK’s work at Los Alamos is supported in part by the U.S. Department
of Energy. P.S. acknowledges financial support by the Deutsche Forschungsgemeinschaft through the project Schm 885/26-1.
\end{acknowledgments}


\begin{thebibliography}{99}

\bibitem{book2a} C. J. Pethick and H. Smith, \textit{Bose-Einstein
condensation in dilute gases} (Cambridge University Press, Cambridge, 2002).

\bibitem{book2} L. P. Pitaevskii and S. Stringari, \textit{Bose-Einstein
Condensation} (Oxford University Press, Oxford, 2003).

\bibitem{emergent} P. G. Kevrekidis, D. J. Frantzeskakis, and R.
Carretero-Gonz\'alez, \textit{Emergent Nonlinear Phenomena in Bose-Einstein
Condensates: Theory and Experiment} (Springer-Verlag, Heidelberg, 2008).

\bibitem{revnonlin} R. Carretero-Gonz\'alez, D. J. Frantzeskakis, and P. G.
Kevrekidis, Nonlinearity \textbf{21}, R139 (2008).

\bibitem{rab} F. Kh. Abdullaev, A. Gammal, A. M. Kamchatnov, and L. Tomio,
Int. J. Mod. Phys. B \textbf{19}, 3415 (2005).

\bibitem{djf} D. J. Frantzeskakis, J.\ Phys.\ A: Math.\ Theor.\ \textbf{43},
213001 (2010).

\bibitem{yuri} Yu. S. Kivshar and G. P. Agrawal, \textit{Optical solitons:
from fibers to photonic crystals}, Academic Press (San Diego, 2003).

\bibitem{motirev} F. Lederer, G.I. Stegeman, D. N. Christodoulides, G. Assanto, M. Segev, and Y. Silberberg, 
Physics Reports {\bf 463} (2008) 1--126.

\bibitem{manakov} S. V. Manakov, Zh. Eksp. Teor. Fiz. {\bf 65}, 505 (1973) [Sov.
Phys. JETP {\bf 38}, 248 (1974)].

\bibitem{ablowitz} M. J. Ablowitz, B. Prinari and A. D. Trubatch, \textit{%
Discrete and Continuous Nonlinear Schr{\"{o}}dinger Systems}, Cambridge
University Press (Cambridge, 2004).

\bibitem{seg1} Z. Chen, M. Segev, T. H. Coskun, D. N. Christodoulides, Yu.
S. Kivshar, and V. V. Afanasjev,
Opt. Lett. \textbf{21}, 1821 (1996).

\bibitem{seg2} 
E. A. Ostrovskaya, Yu. S. Kivshar, Z. Chen, and M. Segev,
Opt. Lett. \textbf{24}, 327 (1999).

\bibitem{seg3} Z. Chen, M. Segev, T. H. Coskun, D. N. Christodoulides and
Yu. S. Kivshar, J. Opt. Soc. Am. B \textbf{14}, 3066-3077 (1997).


\bibitem{buschanglin} Th. Busch and J. R. Anglin, Phys. Rev. Lett. \textbf{87%
}, 010401 (2001).

\bibitem{DDB} H. E. Nistazakis, D. J. Frantzeskakis, P. G. Kevrekidis, B. A.
Malomed, and R. Carretero-Gonz{\'a}lez, Phys. Rev. A \textbf{77}, 033612
(2008).

\bibitem{kanna} M. Vijayajayanthi, T. Kanna, and M. Lakshmanan, Phys. Rev. A
\textbf{77}, 013820 (2008).

\bibitem{rajendran} S. Rajendran, P. Muruganandam, M. Lakshmanan, J. Phys. B
\textbf{42}, 145307 (2009).

\bibitem{val} V. A. Brazhnyi and V. M. P{\'e}rez-Garc{\'i}a, Chaos, Solitons
$\&$ Fractals \textbf{44}, 381 (2011).

\bibitem{berloff} C. Yin, N. G. Berloff, V. M. Perez-Garcia, V. A. Brazhnyi,
and H. Michinel, Phys. Rev. A \textbf{83}, 051605 (2011). 

\bibitem{VB} K. J. H. Law, P. G. Kevrekidis, and L. S. Tuckerman, Phys. Rev.
Lett. \textbf{105}, 160405 (2010).

\bibitem{Alvarez} A. \'Alvarez, J. Cuevas, F. R. Romero, and P. G.
Kevrekidis, Physica D \textbf{240}, 767 (2011).

\bibitem{vaspra} V. Achilleos, P. G. Kevrekidis, V. M. Rothos, and D. J.
Frantzeskakis, Phys. Rev. A \textbf{84}, 053626 (2011).

\bibitem{vasnjp} V. Achilleos, D. Yan, P. G. Kevrekidis, and D. J.
Frantzeskakis, New J. Phys. \textbf{14}, 055006 (2012).

\bibitem{fot} F. Tsitoura, V. Achilleos, B. A. Malomed, D. Yan, P. G. Kevrekidis, 
and D. J. Frantzeskakis, Phys. Rev. A {\bf 87}, 063624 (2013).

\bibitem{yan} D. Yan, F. Tsitoura, P.G. Kevrekidis, 
D.J. Frantzeskakis, arXiv:1402.1895.

\bibitem{sengdb} C. Becker, S. Stellmer, P. Soltan-Panahi, S. D\"{o}rscher,
M. Baumert, E.-M. Richter, J. Kronj\"{a}ger, K. Bongs, and K. Sengstock,
Nature Phys. \textbf{4}, 496 (2008).

\bibitem{peter1} S. Middelkamp, J. J. Chang, C. Hamner, R.
Carretero-Gonz\'alez, P. G. Kevrekidis, V. Achilleos, D. J. Frantzeskakis,
P. Schmelcher, and P. Engels, Phys. Lett. A \textbf{375}, 642 (2011).

\bibitem{peterprl} C. Hamner, J. J. Chang, P. Engels, M. A. Hoefer, Phys.
Rev. Lett. \textbf{106}, 065302 (2011).

\bibitem{peter2} D. Yan, J. J. Chang, C. Hamner, P. G. Kevrekidis, P.
Engels, V. Achilleos, D. J. Frantzeskakis, R. Carretero-Gonz\'alez, and P.
Schmelcher, Phys. Rev. A \textbf{84}, 053630 (2011).

\bibitem{peterpra} M. A. Hoefer, J. J. Chang, C. Hamner, and P. Engels,
Phys. Rev. A \textbf{84}, 041605 (2011).

\bibitem{peter3} D. Yan, J. J. Chang, C. Hamner, M. A. Hoefer, P. G.
Kevrekidis, P. Engels, V. Achilleos, D. J. Frantzeskakis, and J. Cuevas, J.
Phys. B \textbf{45}, 115301 (2012).

\bibitem{skryabin} D. V. Skryabin, Phys. Rev. A {\bf 63}, 013602 (2000).

\bibitem{kody} K. J. H. Law, P. G. Kevrekidis, and L. S. Tuckerman, Phys. Rev.
Lett. {\bf 105}, 160405 (2010); see also {\bf 106}, 199903(E) (2011).

\bibitem{andersonexp} B. P. Anderson, P. C. Haljan, C. E. Wieman, and E. A. Cornell,
Phys. Rev. Lett. {\bf 85}, 2857 (2000).

\bibitem{pola}  M. Pola, J. Stockhofe, P. Schmelcher,
P.G. Kevrekidis, Phys. Rev. A {\bf 86}, 053601 (2012).


\bibitem{Morsch} O. Morsch and M. Oberthaler, 
Rev. Mod. Phys. \textbf{78},
(2006) 179--215.


\bibitem{smerzi} S. Raghavan, A. Smerzi, S. Fantoni, and S. R. Shenoy, 
Phys.
Rev. A \textbf{59} (1999) 620--633. 

\bibitem{smerzi1} S. Raghavan, A. Smerzi, and V. M. Kenkre,
Phys. Rev. A \textbf{60} (1999) R1787-R1790. 

\bibitem{smerzi2} A. Smerzi and S. Raghavan, 
Phys.
Rev. A \textbf{61} (2000) 063601.

\bibitem{markus1} M. Albiez, R. Gati, J. F\"{o}lling, S. Hunsmann, M.
Cristiani, and M. K. Oberthaler, 
Phys. Rev. Lett. \textbf{95} (2005) 010402.

\bibitem{kiv2} E. A. Ostrovskaya, Yu. S. Kivshar, M. Lisak, B. Hall, F.
Cattani, and D. Anderson, 
Phys. Rev. A \textbf{61} (2000) 031601(R).

\bibitem{mahmud} K. W. Mahmud, J. N. Kutz, and W. P. Reinhardt, 
Phys. Rev. A
\textbf{66} (2002) 063607.

\bibitem{bam} V. S. Shchesnovich, B. A. Malomed, and R. A. Kraenkel, 
Physica
D \textbf{188} (2004) 213--240.

\bibitem{Bergeman_2mode} D. Ananikian and T. Bergeman, 
Phys. Rev. A \textbf{%
73} (2006) 013604.

\bibitem{infeld2} P. Zi\'{n}, E. Infeld, M. Matuszewski, G. Rowlands, and M.
Trippenbach, 
Phys. Rev. A \textbf{73} (2005) 022105.

\bibitem{todd} T. Kapitula and P. G. Kevrekidis, 
Nonlinearity \textbf{18} (2005)
2491--2512.

\bibitem{theo} G. Theocharis, P. G. Kevrekidis, D. J. Frantzeskakis, and P.
Schmelcher, 
Phys. Rev. E \textbf{74} (2006) 056608.

\bibitem{carr} D. R. Dounas-Frazer, A. M. Hermundstad, and L. D. Carr, 
Phys.
Rev. Lett. \textbf{99} (2007) 200402.

\bibitem{julia} B. Juli{\'a}-D{\'i}az, D. Dagnino, M. Lewenstein,
J. Martorell and A. Polls,
Phys. Rev. A {\bf 81}, 023615 (2010) 023615.

\bibitem{bettina} B. Gertjerenken and C. Weiss,
Phys. Rev. A {\bf 88} (2013) 033608.

\bibitem{julia2} B. Juli{\'a}-D{\'i}az, T. Zibold, M.K. Oberthaler,
M. Mel{\'e}-Messeguer, J. Martorell and A. Polls,
Phys. Rev. A {\bf 86} (2012) 023615.



\bibitem{pseudo} T. Mayteevarunyoo, B. A. Malomed, and G. Dong, 
Phys. Rev. A
\textbf{78} (2008) 053601.

\bibitem{fibers} C. Par\'{e} and M. Florja\'{n}czyk, 
Phys. Rev. A \textbf{41}%
(1990) 6287--6295.

\bibitem{HaeltermannPRL02} C. Cambournac, T. Sylvestre, H. Maillotte , B.
Vanderlinden, P. Kockaert, Ph. Emplit, and M. Haelterman, 
Phys. Rev. Lett.
\textbf{89} (2002) 083901.

\bibitem{zhigang} P.G. Kevrekidis, Z. Chen, B. A. Malomed, D. J.
Frantzeskakis, and M. I. Weinstein,
 Phys. Lett. A \textbf{340} (2005) 275--280.

\bibitem{boris_book} B.A. Malomed (Ed.),
{\it Spontaneous symmetry-breaking, self-trapping and Josephson
oscillations}, Springer-Verlag (Heidelberg, 2013).

\bibitem{zibold} T. Zibold, E. Nicklas, C. Gross
and M.K. Oberthaler,
Phys. Rev. Lett. {\bf 105} (2010) 204101. 

\bibitem{leblanc} L. J. LeBlanc, A. B. Bardon, J. McKeever, M. H. T. Extavour, D. Jervis, J. H. Thywissen, F. Piazza, and A. Smerzi,
Phys. Rev. Lett. {\bf 106} (2011) 025302.

\bibitem{watanabe1} G. Watanabe and H. M\"akel\"a,
Phys. Rev. A {\bf 85}, 053624 (2012).

\bibitem{strzys} M. P. Strzys, E. M. Graefe, and H. J. Korsch,
New J. Phys. {\bf 10} (2008) 013024.

\bibitem{abdullaev} F. Kh. Abdullaev and R. A. Kraenkel,
Phys. Rev. A {\bf 62}, 023613 (2000).

\bibitem{boukobza} E. Boukobza, M. G. Moore, D. Cohen, and A. Vardi,
Phys. Rev. Lett. {\bf 104} (2010) 240402.

\bibitem{watanabe2} G. Watanabe, 
Phys. Rev. A {\bf 81}, 021604(R) (2010).

\bibitem{newtonsolver} C. T. Kelley, \textit{Solving Nonlinear Equations with Newton's Method} (Society for Industrial and Applied Mathematics, Philadelphia, 2003).

\bibitem{chaostodd} T. Kapitula and P.G. Kevrekidis,
Chaos {\bf 15}, 037114, 13 pages (2005).

\bibitem{tsubota} M. Eto, K. Kasamatsu, M. Nitta, H. Takeuchi, and 
M. Tsubota,
Phys. Rev. A {\bf 83}, 063603 (2011).



\end{thebibliography}
\end{document}